%% file: main-final.tex
\documentclass[runningheads]{llncs}
\usepackage{graphicx}
\usepackage{amsmath,amssymb}
\usepackage{enumitem}
\usepackage{bm}
\usepackage{mathdots}
\usepackage{mathtools}
\usepackage{algorithmic}
\usepackage{soul} 
\usepackage{comment}
\usepackage{letltxmacro}
\usepackage{nameref}
\usepackage[X2,T2A]{fontenc}

\input{mymacros.tex}

\usepackage{tikz}

\usepackage[textsize=tiny]{todonotes}

\newcommand{\nn}{{n}}   

\begin{document}
\title{Discovering Polynomial and Quadratic Structure in Nonlinear Ordinary Differential Equations}
\titlerunning{Discovering Polynomial and Quadratic Structure in Nonlinear ODEs}
%
\author{Boris Kramer\inst{1}\orcidID{0000-0002-3626-7925} \and
Gleb Pogudin\inst{2}\orcidID{0000-0002-5731-8242}}
\authorrunning{B. Kramer and G. Pogudin}
%
\institute{University of California San Diego, San Diego, California, U.S.A \\
\email{bmkramer@ucsd.edu} 
\and
LIX, CNRS, \'Ecole Polytechnique, Institute Polytechnique de Paris, France
\email{gleb.pogudin@polytechnique.edu} }
\maketitle              
\begin{abstract}
Dynamical systems with quadratic or polynomial drift exhibit complex dynamics, yet compared to nonlinear systems in general form, are often easier to analyze, simulate, control, and learn.
Results going back over a century have shown that the majority of nonpolynomial nonlinear systems can be recast in polynomial form, and their degree can be reduced further to quadratic. 
This process of polynomialization/quadratization reveals new variables (in most cases, additional variables have to be added to achieve this) in which the system dynamics adhere to that specific form, which leads us to discover new  structures of a model.
This chapter summarizes the state of the art for the discovery of polynomial and quadratic representations of finite-dimensional dynamical systems. We review known existence results, discuss the two prevalent algorithms for automating the discovery process, and give examples in form of a single-layer neural network and a phenomenological model of cell signaling.
\end{abstract}
%
%

\section{Introduction}

\subsection{The benefits of specific model structure}
Discovering and then leveraging structure in the governing equations of scientific and engineering processes can reveal important fundamental mechanisms, and simplify simulation, control, optimization, and model learning, among many others. In this chapter, we focus on the symbolic form of the governing equations, and discuss theoretical results, methods and algorithms to discover a quadratic or polynomial structure of a given system model. 
Evolutionary processes in engineering and science are often modeled with nonlinear non-autonomous ordinary or partial differential equations (ODEs/PDEs) that describe the time evolution of the states of the system, i.e., the physically necessary and relevant variables. However, these models are not unique: the same evolutionary process can be modeled with different variables, resulting in a different symbolic structure of the equations.
This idea of variable transformation (referred to as \textit{lifting} when extra variables are added) to promote model structure is found across different communities, with literature spanning over a century. The benefits of variable transformations are wide-ranging.
%
%
In fluid dynamics, variable transformations have been exploited to guarantee stability~\cite{hughes1986new,kalashnikova2011stable,rezaian2020impact}, conservation principles of fluid and plasma equations~\cite{halpern_2021_antisymmetric}, as well as to uncover a linear structure of the Burgers' PDE through the famous Cole-Hopf transformation~\cite{cole1951quasi,hopf1950partial}. 
In the field of dynamical systems, the dynamic mode decomposition~\cite{rowley2009spectral,schmid2010dynamic} computes part of the spectrum of the Koopman operator; it has been shown in \cite{williams2015data,netto2021analytical} that the choice of variables to learn that spectrum from is critical to its success.
In control engineering, feedback linearization is a process that transforms a nonlinear system into a structured linear model \cite{jakubczyk1980linearization,Khalil_NonlinearSystems} through a specific nonlinear state transformation. The transformed state can be augmented; that is, it can have an increased dimension relative to the original state. 
A change of variables can also ensure that physical constraints are more easily met in a simulation~\cite{nam2011space,hassler2020finite}, it can improve analysis~\cite{savageau1987recasting,liu2015abstraction,Brenig2018}, and the resulting transformed systems can be solved faster numerically~\cite{savageau1987recasting}.

\subsection{Leveraging polynomial and quadratic model structure}
The discovery of a specific model structure, such as polynomial or quadratic, has thus seen broad interest. 
In the context of optimization, McCormick~\cite{mccormick1976computability} used variable substitutions to achieve a quadratic structure so that non-convex optimization problems can be recast as convex problems in the new variables.
In fluid mechanics, the quadratic model structure of the specific volume variable representation has been exploited in~\cite{balajewicz2016minimal} to allow model stabilization.
To analyze equilibrium branches of geometrically nonlinear finite element models, the authors in \cite{guillot2019generic} recast the model into a quadratic form, for which the Jacobian and a specific Taylor series can be easily obtained. Then, they use the Asymptotic Numerical Method to find the equilibrium branches.
One can also use a lifted quadratic system in order to facilitate reachability analysis of the model~\cite{Forets_2021}.
In the area of analog computing with chemical reaction networks, quadratic forms represent elementary chemical reactions. Therefore, transforming a polynomial ODE into a quadratic one can be used to establish the Turing completeness of elementary chemical reactions~\cite{bournez2007polynomial,fages2017strong,lifeware1}, see also Section~\ref{ss:sigmoid} below.
Moreover, quadratic-bilinear model structure is appealing for intrusive model reduction
~\cite{gu2011qlmor,bennerBreiten2015twoSided,bennergoyal2016QBIRKA,KW18nonlinearMORliftingPOD,KW2019_balanced_truncation_lifted_QB}, as it eliminates the need for additional hyperreduction/interpolation to reduce the expense of evaluating the nonlinear term~\cite{deim2010,astrid2008missing,barrault2004empirical,carlberg2013gnat,nguyen2008best}, where in some cases the number of interpolation points required for accuracy eliminates computational gains~\cite{bergmann2009enablers,huangAIAA18RomRocketCombustion}.
The \textit{Lift \& Learn} method \cite{QKPW2020_lift_and_learn,QKMW2019_transform_and_learn} and related work~\cite{SKHW2020_learning_ROMs_combustor,gosea2018data,JQK2021_performanceCompCombustion,McQuarrie_regularizedOPINF} leverage lifting transformations to learn low-order polynomial ROMs of complex nonlinear systems, such as combustion dynamics, from lifted data. For quadratic and cubic model structures one can equip these learned ROMs with stability guarantees; see~\cite{K2020_stability_domains_QBROMs,SKP_PIregulartizationOPINF}.

\subsection{Outline of this chapter}
The goal of this chapter is to familiarize the reader with the concept of discovering a symbolic structure through polynomialization and quadratization, existing results thereof, and algorithms and software that turn theoretical results into practical use. In Section~\ref{sec:Polynomialization} we focus on the polynomialization problem for autonomous ODEs, discuss classes of ODEs for which existence of a polynomialization can be shown, and present two software packages that can (suboptimally) compute those. 
In Section~\ref{sec:Quadratization} we follow a similar structure for the quadratization problem. Section~\ref{sec:application} then presents two representative applications in the form of a single-layer neural network and a phenomenological model of cell signaling. In Section~\ref{sec:problems} we outline several open problems in the field, and in Section~\ref{sec:conclusions} we conclude our findings.

\subsection{Notation and definitions} \label{ss:notation}
The set of nonnegative integers is denoted as $\mathbb{Z}_{\geqslant 0}$.
Throughout the paper, we will use $\KK$ to denote the base field which can be $\real$ or $\compl$.
We denote by $\bx= [x_1, x_2, \ldots, x_\nn]^\top$ an $\nn$-dimensional column vector in $\KK^n$, which generically represents the state of a dynamical system and with $\dot{\bx}$ its derivative. Moreover, we denote by $\bu=\bu(t) \in \KK^r$ a generic input vector. We often omit the explicit dependence of $t$ for ease of notation. 
%

%
%
%

A product of positive-integer powers of variables is referred to as \textit{monomial} (e.g., $x^5y$) and the total degree of a monomial is the sum of the powers of the variables appearing in it. A \textit{polynomial} is a sum of monomials, e.g., $x + y^2$.
By $\KK[\bx]$ and $\KK[\bx, \bw]$ we denote the sets of all polynomials with (possibly complex) coefficients in $\bx$ and $\bx, \bw$, respectively. 
The sets $\KK[\bx,\bw,\bu]$ and $\KK[\bx,\bw,\bu, \dot{\bu}]$ are sets of polynomials defined similarly.
Let $p(\bx)$ be a polynomial, then $\deg_{x_i} p(\bx)$ denotes the degree of $p$ with respect to $x_i$, that is, the maximal power of $x_i$ appearing in $p(\bx)$. Similarly, $\deg p(\bx)$  denotes the total degree of $p$, that is, the maximum of the total degrees of the monomials appearing in $p(\bx)$.
For example $\deg_x (x^5y) = 5$ and $\deg (x^5 y) = 6$.
The degree of a vector or matrix is defined as the maximum of the degrees of its entries.

\section{Polynomialization} \label{sec:Polynomialization}
We formulate the polynomialization problem in Section~\ref{ss:poly:formulation}. Then, in Section~\ref{ss:poly:theoretical} we summarize existing theoretical results about polynomialization of ODEs, and in Section~\ref{ss:poly:algorithmic} we turn to the practical problem of how to find a polynomialization algorithmically, and how existing algorithms differ in their approach.

\subsection{Problem formulation} \label{ss:poly:formulation}

\begin{definition}[Polynomialization]\label{def:poly}
  Consider a system of ODEs
  \begin{equation}\label{eq:sys_main_poly}
  \dot{\bx} = \bbf(\bx),
  \end{equation}
  where $\bx = \bx(t) = [x_1(t), \ldots, x_\nn(t)]^\top$ and 
  $\bbf(\bx) = [f_1(\bx), \ldots, f_\nn(\bx)]^\top$ is a vector of $\KK$-valued functions on a domain $\mathcal{D} \subset \KK^\nn$.
  Then a vector of differentiable functions $\bw = [w_1(\bx), \ldots, w_\ell(\bx)]^\top$ defined on $\mathcal{D} \subset \KK^\nn$
  is said to be a \emph{polynomialization} of~\eqref{eq:sys_main_poly} if there exist vectors $\bq_1(\bx, \bw)$ and $\bq_2(\bx, \bw)$ of polynomials of dimensions $\nn$ and $\ell$, respectively such that 
  \begin{equation}
  \dot{\bx} = \bq_1(\bx, \bw) \quad \text{ and }\quad \dot{\bw} = \bq_2(\bx, \bw)
  \end{equation}
  for every $\bx$ solving~\eqref{eq:sys_main_poly}.
  The dimension $\ell$ of the vector $\bw$ is called the \emph{order of polynomialization}.
  A polynomialization of the smallest possible order is called an \emph{optimal polynomialization}.
\end{definition}

One can rephrase the definition above saying that the map $\bx \to [\bx, \bw]$ sends the trajectories of the original system to trajectories of an ODE system defined by polynomial right-hand sides.

\begin{example}
    Let $\KK = \compl$, $\nn = 1$, $\mathcal{D} = \compl$, and consider a scalar ODE $\dot{x} = e^{-ax}$.
    We claim that $w(x) = e^{-ax}$ is a polynomialization of the ODE.
    Since it includes only one additional variable, it is clearly optimal.
    Indeed, we obtain:
    \[
      \dot{x} = e^{-ax} = w\quad\text{ and }\quad \dot{w} = -a\dot{x}e^{-ax} = -ae^{-2ax} = -aw^2,
    \]
    so we can take $q_1(x, w) = w$ and $q_2(x, w) = -aw^2$.
\end{example}

\begin{example}
    Let $\KK = \real$, $\nn = 1$, $\mathcal{D} = \{x \in \real \colon x > 0\}$, and consider a scalar ODE $\dot{x} = x^a$.
    We claim that $\bw = [x^a,\; 1/x]^\top$ is a polynomialization of the ODE.
    To check this, we compute:
    \[
    \dot{x} = x^a = w_1, \quad \dot{w}_1 = a\dot{x}x^{a - 1} = a w_1^2 w_2, \quad \dot{w}_2 = -\frac{\dot{x}}{x^2} = -w_1 w_2^2,
    \]
    so we can take $q_1(x, \bw) = w_1$ and $\bq_2(x, \bw) = [a w_1^2 w_2,\; -w_1 w_2^2]^\top$.
\end{example}

\subsection{Overview of theoretical results} \label{ss:poly:theoretical}

The existence of a polynomialization for a system with non-polynomial right-hand side has been rediscovered several times in the literature. The main difference between the existing results is the class of functions $\bbf$ that are considered for the right-hand side of~\eqref{eq:sys_main_poly}.
Those may be polynomials with non-integer degrees~\cite{carravetta2020solution} or a set of functions obtained from a given list of ``basis'' functions (typically including $e^x, \sin x, \ln x$) using arithmetic operations and compositions~\cite{gu2011qlmor,lifeware2,savageau1987recasting}. 
It has been pointed out several times~\cite{kerner1981universal,HernndezBermejo1997,lifeware2} that there is flexibility in choosing the set of ``basis'' function as long as the chosen functions themselves ``come from a differential equation''.
We will make this precise below by using the notion of a differentially-algebraic function.

\begin{definition}[Differentially-algebraic function]
   Consider a domain $\mathcal{D} \subset \KK^\nn$ in a space with coordinates $\bx$.
   An $\nn$-variate analytic function $f \colon \mathcal{D} \to \KK$ is called \emph{differentially-algebraic} if, for every $1 \leqslant i \leqslant \nn$, $f$ satisfies a polynomial differential equation with respect to $x_i$, that is, there exists and integer $h$ and a polynomial $P$ in $h + 1$ variables such that
   \[
   P\left(f, \frac{\partial f}{\partial x_i}, \ldots, \frac{\partial^{h} f}{\partial x_i^{h}}\right) = 0.
   \]
\end{definition}

Many of the functions used in modeling processes in science and engineering are differentially-algebraic. In particular, $e^x$ (satisfies $f' = f$), $\sin x$ (satisfies $f'' + f = 0$), and $\ln x$ (satisfies $f'' = -2 (f')^2$). The following century-old proposition allows one to verify differential algebraicity for a large class of functions.

\begin{proposition}[{\cite[\S 6, Satz 17]{Ostrowski1920}\footnote{An English translation of the German paper from 1920 is available at~\cite{hansen2022alexander}.}}]\label{prop:diffalg_properties}
    \begin{enumerate}
        \item[]
        \item Consider a domain $\mathcal{D} \subset \KK^\nn$ and differentially-algebraic functions $f_1, f_2$ defined on $\mathcal{D}$.
        Then $f_1 + f_2$ and $f_1 f_2$ are differentially algebraic as well.
        Furthermore, if $f_2$ does not vanish on $\mathcal{D}$, then $\frac{f_1}{f_2}$ is differentially algebraic.

        \item Let $f_1, \ldots, f_{r}$ be differentially-algebraic functions on a domain $\mathcal{D} \subset \KK^n$ and $\mathcal{U} \subset \KK^r$ be a domain containing the range of the map 
        \[
        \bx \to [f_1(\bx), \ldots, f_r(\bx)]^\top.
        \]
        Then, for a differentially-algebraic function $g$ on $\mathcal{U}$, the composition
        \[
        g(f_1(\bx), \ldots, f_r(\bx))
        \]
        is differentially-algebraic on $\mathcal{D}$.
    \end{enumerate}
\end{proposition}

\noindent 
We can use Proposition~\ref{prop:diffalg_properties} to show that the function 
   \[
   f(x_1, x_2) = \frac{\ln(1 + x_1^2 + x_2^4)}{e^{x_1} + e^{x_1 x_2}}
   \]
is differentially-algebraic. First, we note that $1 + x_1^2 + x_2^4$ and $x_1x_2$, as any other polynomials, are differentially-algebraic since their high-order derivatives vanish. Next, the functions $e^x$ and $\ln(x)$ are differentially-algebraic as explained above. 
Using the second part of Proposition~\ref{prop:diffalg_properties}, we deduce that the compositions $\ln(1 + x_1^2 + x_2^4),\; e^{x_1}, \; e^{x_1 x_2}$ are differentially-algebraic. Finally, since $f(x_1, x_2)$ can be obtained from these functions by arithmetic operations, we establish its differential algebraicity using the first part of Proposition~\ref{prop:diffalg_properties}.

Having defined differentially-algebraic functions, we next state a general theorem for the existence of polynomialization, as shown in~\cite[Section~2]{HernndezBermejo1995}.

\begin{theorem}[Existence of polynomialization]\label{thm:polynomialization}
    Consider the system~\eqref{eq:sys_main_poly}.
    If the functions $\bbf$ are differentially-algebraic with respect to $\mathbf{x}$ on $\mathcal{D}$, then there exists a polynomialization for~\eqref{eq:sys_main_poly}.
\end{theorem}

Since any function obtained from elementary functions such as $e^x, \sin x, \ln x$, and others, is differentially-algebraic when arithmetic operations and compositions are applied, Theorem~\ref{thm:polynomialization} encompasses all the polynomialization existence results we are aware of, such as~\cite{kerner1981universal,HernndezBermejo1997,gu2011qlmor,carravetta2020solution,lifeware2}.

\subsection{Overview of algorithmic results and software} \label{ss:poly:algorithmic}
To the best of our knowledge, all the approaches to proving the existence of polynomialization under various assumptions including~\cite{HernndezBermejo1995,lifeware2,kerner1981universal,carravetta2020solution,gu2011qlmor,savageau1987recasting} are constructive and can be turned into algorithms.
Doing this in practice, however, requires resolving certain challenges when implementing those results, such as operating with arbitrary functional expressions, performing differentiation and substitution on them, etc.
We are aware of only two software tools which can perform polynomialization in practice: {\sc BioCham}~\cite{lifeware2} and {\sc QBee}~\cite{QBee,ByIsPoKr_Quadratization_2023}.
Both tools take as input the symbolic representation of the right-hand side of the differential equation, $\bbf(\bx)$, generated from a set of ``basis'' functions using arithmetic operations and compositions. 
Those include exponential, logarithmic, trigonometric, hyperbolic functions, and their inverses.
While this class is sufficient for the majority of the applications, it is only a subset of the large set of differentially-algebraic functions. The latter includes, for example, the Bessel and Airy functions and Painlev\'e transcendents. Thus, not every system that is polynomializable as per Theorem~\ref{thm:polynomialization} can be treated with the two existing software packages, see Example~\ref{ex:bessel}.

In general, both {\sc BioCham} and {\sc QBee} follow the same strategy: at each step, the algorithms take one nonpolynomial term and add it as a new variable.
{\sc BioCham} does this in a systematic fashion taking into account properties of specific functions, which allows it to achieve quadratic time complexity~\cite[Proposition~2]{lifeware2}.
{\sc QBee} instead tries many different substitutions, which may take more time (e.g., there is no good complexity bound) but the produced polynomialization may be of lower order~\cite[Example~5.5]{ByIsPoKr_Quadratization_2023} due to a more comprehensive search.

Developing algorithms and software with such guarantees (even under some restrictions on the right-hand side, e.g., allowing only exponential functions as ``basis'' functions) is an open problem.
We thus point out that neither of the discussed software tools provides optimality guarantees for the produced polynomialization. We illustrate this next on an example. 
 
\begin{example}[Suboptimality of current software]\label{ex:poly_nonoptimal}
    Consider the scalar ODE $\dot{x} = e^{-x} + e^{-1.5x}$.
    Both {\sc BioCham} and {\sc QBee} find a polynomialization of order two with $\bw = [e^{-x}, e^{-1.5 x}]^\top$.
    However, this polynomialization is not optimal, as we find that this equation can be polynomialized with only one new variable $w(x) = e^{-0.5x}$:
    \[
    \begin{cases}
      \dot{x} = w^2 + w^3,\\
      \underline{\dot{w}} = -0.5\cdot\dot{x}\cdot e^{-0.5x} = -0.5(e^{-1.5x} + e^{-2x}) = \underline{-0.5 (w^3 + w^4)}.
    \end{cases}
    \]
\end{example}

\begin{example}[Limitation of the current software]\label{ex:bessel}
    We fix a complex number $\alpha \in \mathbb{C}$ and consider the scalar ODE $\dot{x} = J_{\alpha}(x)$, where $J_{\alpha}$ is the Bessel function of the first kind.
    Neither {\sc QBee} nor {\sc BioCham} can be applied to this equation but it can be polynomialized. 
    Recall, that $J_{\alpha}$ satisfies a differential equation
    \begin{equation}\label{eq:bessel}
      t^2 \ddot{J}_{\alpha} + t \dot{J}_{\alpha} + (t^2 -\alpha) J_{\alpha} = 0.
    \end{equation}
    We introduce new variables $\bw = [1/t, J_{\alpha}, \dot{J}_{\alpha}]$ and use~\eqref{eq:bessel} to find the polynomial ODE system
    \[
    \begin{cases}
        \dot{x} = J_{\alpha} = w_2,\\
        \dot{w}_1 = \frac{-1}{t^2} = -w_1^2,\\
        \dot{w}_2 = \dot{J}_{\alpha} = w_3,\\
        \dot{w}_3 = \ddot{J}_{\alpha} = -\frac{\dot{J}_{\alpha}}{t} + (\frac{\alpha}{t^2} - 1) J_{\alpha} = -w_1w_3 + (\alpha w_1^2 - 1)w_2.
    \end{cases}
    \]
    Thus, a natural direction for improving the existing software would be to take as input user-defined differential algebraic functions (e.g., $J_{\alpha}$) described by their differential equations (as~\eqref{eq:bessel} in this example).
\end{example}

\section{Quadratization} \label{sec:Quadratization}
We formulate the quadratization problem in Section~\ref{ss:quad:formulation}. Then, in Section~\ref{ss:quad:theory} we summarize existing theoretical results about quadratization of ODEs, both autonomous and non-autonomous. In Section~\ref{ss:quad:algorithms} we survey existing algorithms to find an (optimal) quadratization, and software implementations thereof. 

\subsection{Problem formulation} \label{ss:quad:formulation}
\begin{definition}[Quadratization]\label{def:quadr}
  Consider a polynomial system of ODEs
  \begin{equation}\label{eq:sys_main}
  \dot{\bx} = \bp(\bx),
  \end{equation}
  where $\bp(\bx) = [p_1(\bx), \ldots, p_\nn(\bx)]^\top$ with $p_1, \ldots, p_\nn \in \mathbb{C}[\bx]$. Then an $\ell$-dimensional vector of new variables
  \begin{equation}\label{eq:quadr}
      \bw = \bw(\bx) \quad \in \ \mathbb{C}[\bx]^\ell
  \end{equation}
  is said to be a \emph{quadratization} of~\eqref{eq:sys_main} if there exist vectors $\bq_1(\bx, \bw)$ and $\bq_2(\bx, \bw)$ of dimensions $\nn$ and $\ell$, respectively, with the entries being polynomials of total degree at most two such that 
  \begin{equation}
  \dot{\bx} = \bq_1(\bx, \bw) \quad \text{ and }\quad \dot{\bw} = \bq_2(\bx, \bw).
  \end{equation}
  The dimension $\ell$ of vector $\bw$ is called the \emph{order of quadratization}.
  A quadratization of the smallest possible order is called an \emph{optimal quadratization}.
  If all the polynomials $w_1(\bx), \ldots, w_\ell(\bx)$ are monomials, the quadratization is called \emph{a monomial quadratization}.
  If a monomial quadratization of a system has the smallest possible order among all the monomial quadratizations of the system, it is called \emph{an optimal monomial quadratization}.
\end{definition}

Similarly to polynomialization, one can think of quadratization as a map $\bx \to  [\bx, \bw]$ sending the trajectories of the system to trajectories of an ODE system with at most quadratic nonlinearities.

\begin{example}\label{ex:toy_quadr}
    Consider the quartic ODE $\dot{x} = x^4$.
    We claim that $w(x) = x^3$ is a (optimal) quadratization of the ODE, which we verify by
    \[
      \dot{x} = xw \qquad \text{and} \qquad \dot{w} = 3\dot{x}x^2 = 3w^2,
    \]
    so we can take $q_1(x, w) = xw$ and $q_2(x, w) = 3w^2$.
\end{example}

\begin{definition}[Quadratization of non-autonomous polynomial ODEs]\label{def:quadr_input}
  Consider a polynomial system of ODEs with external inputs,
  \begin{equation}\label{eq:sys_main_input}
  \dot{\bx} = \bp(\bx, \bu),
  \end{equation}
  where $\bx = \bx(t) = [x_1(t), \ldots, x_\nn(t)]^\top$ are the states, $\bu = \bu(t) = [u_1(t), \ldots, u_r(t)]^\top$ denote the inputs and 
  $p_1, \ldots, p_\nn \in \mathbb{C}[\bx, \bu]$.
  Then an $\ell$-dimensional vector of new variables
  \begin{equation}\label{eq:quadr_input}
      \bw = \bw(\bx, \bu) \quad \in \ \mathbb{C}[\bx, \bu]^\ell
  \end{equation}
 is said to be a \emph{quadratization} of~\eqref{eq:sys_main_input} if there exist vectors $\bq_1(\bx, \bw, \bu, \dot{\bu})$ and $\bq_2(\bx, \bw, \bu, \dot{\bu})$ of dimensions $\nn$ and $\ell$, respectively, with the entries being polynomials of total degree at most two such that 
  \begin{equation}
  \dot{\bx} = \bq_1(\bx, \bw, \bu, \dot{\bu}) \quad \text{ and }\quad \dot{\bw} = \bq_2(\bx, \bw, \bu, \dot{\bu}).
  \end{equation}
  The number $\ell$ is called the \emph{order of quadratization}.
  A quadratization of the smallest possible order is called an \emph{optimal quadratization}. 
  A \emph{monomial quadratization} of non-autonomous systems is defined similarly as in Definition~\ref{def:quadr}.
\end{definition}

It may happen that the input function $\bu(t)$ is not differentiable. 
While it is not always possible to relax the differentiability condition on the inputs, we characterize the cases when it is possible. 

\begin{definition}[Input-free quadratization of non-autonomous polynomial ODEs]\label{def:quadr_input_zero}
  In the notation of Definition~\ref{def:quadr_input}, an $\ell$-dimensional vector of new variables 
  \begin{equation}
      \bw = \bw(\bx) \quad \in \ \mathbb{C}[\bx]^\ell
  \end{equation}
  is said to be an \emph{input-free quadratization} of~\eqref{eq:sys_main_input} 
  if there exist vectors $\bq_1(\bx, \bw, \bu)$ and $\bq_2(\bx, \bw, \bu)$ of dimensions $\nn$ and $\ell$, respectively, with the entries being polynomials of total degree at most two such that 
  \begin{equation}\label{eq:input_free_result}
  \dot{\bx} = \bq_1(\bx, \bw, \bu) \quad \text{ and }\quad \dot{\bw} = \bq_2(\bx, \bw, \bu).
  \end{equation}
\end{definition}

\subsection{Overview of theoretical results} \label{ss:quad:theory}

We start by stating an existence result that has been rediscovered many times (as early as 1902), and in different contexts, see~\cite{Appelroth1902,Lagutinskii,kerner1981universal,Kinyon1995,gu2011qlmor,carravetta2015global,Brenig2018,Ohtsuka}. The version we provide below is from~\cite[Theorem~1]{CPSW05}.

\begin{theorem}\label{thm:existence_simple}
  For every ODE system of the form~\eqref{eq:sys_main}, there exists a monomial quadratization.
  Furthermore, if $d_i := \deg_{x_i}\bp(\bx)$, then the order of an optimal monomial quadratization does not exceed $\prod_{i=1}^\nn (d_i + 1)$.
\end{theorem}

We give a proof of this result below, since the ideas from it have been used both in further extensions to non-autonomous systems and in the algorithm designs that we discuss in Section~\ref{ss:quad:algorithms}.

\begin{proof}[of Theorem~\ref{thm:existence_simple}]
    We consider the set of monomials 
    \[
    \mathcal{M} := \{x_1^{e_1}\cdots x_\nn^{e_\nn} \ \mid \ \forall\;  1 \leqslant i \leqslant \nn \colon 0 \leqslant e_i \leqslant d_i,\; e_i \in \mathbb{Z}\}.
    \]
    We claim that taking $\bw(\bx)$ to be a vector consisting of all the elements of $\mathcal{M}$ will yield a quadratization for the original system~\eqref{eq:sys_main}.
    Since every monomial in the right-hand side of the original system, $\bp(\bx)$, belongs to $\mathcal{M}$, $\bq_1$ can be chosen to be a linear polynomial.
    Now we consider an element $m \in \mathcal{M}$. 
    We can write its derivative as
    \[
    \dot{m} = \sum\limits_{i = 1}^\nn p_i(\bx) \frac{\partial m}{\partial x_i}.
    \]
    Consider any $i$ with $\frac{\partial m}{\partial x_i} \neq 0$.
    Then $\frac{\partial m}{\partial x_i}$ is of the form $c \frac{m}{x_i}$ for some constant $c$, where $\frac{m}{x_i} \in \mathcal{M}$.
    Furthermore, every monomial appearing in $p_i(\bx)$ belongs to $\mathcal{M}$ as well.
    Therefore, $p_i(\bx) \frac{\partial m}{\partial x_i}$ is a quadratic polynomial in the elements of $\mathcal{M}$.
    Thus, $\mathcal{M}$ indeed provides a quadratization.
    We conclude the proof by observing that $|\mathcal{M}| = \prod\limits_{i = 1}^\nn (d_i + 1)$.\qed 
\end{proof}

The given proof is constructive but the resulting algorithm is not practical as it introduces $\prod_{i=1}^\nn (d_i + 1)$ new variables, and this number is large (e.g., exponential in the dimension of the system). 
For instance, in Example~\ref{ex:toy_quadr} the construction from the proof above would introduce three new variables $x^2, x^3, x^4$ instead of just one.
We are not aware of any tighter upper bounds for the number of new variables. 
Furthermore, it has been conjectured~\cite[Conjecture~1]{lifeware1} that there exist systems requiring exponentially many new variables.
While the conjecture is still open, it has been shown~\cite[Theorem~2]{lifeware1} that the problem of finding an optimal quadratization is NP-hard.
Of note, if one allows the new variables ($\bw$ in terms of Definition~\ref{def:quadr}) to be Laurent monomials (that is, monomials with arbitrary integer powers), then a much better bound is available, as we see next.

\begin{theorem}[{\cite[Proposition~1]{Bychkov2021}}]\label{thm:bound_laurent}
  For every ODE system of the form~\eqref{eq:sys_main}, there exists a Laurent monomial quadratization with the order not exceeding the total number of terms in the right-hand side (that is, in $\bp$).
\end{theorem}
 
In engineering and science, systems are often forced externally by disturbances, control inputs, or through coupling to other subsystems. For such non-autonomous ODEs, it can also be shown that a quadratization exists. 

\begin{theorem}[{\cite[Theorem~3.2]{ByIsPoKr_Quadratization_2023}}]\label{thm:input_quadr}
  Consider the polynomial ODE system with external inputs $\bu(t)$ of the form
  \begin{equation}
  \dot{\bx} = \bp(\bx, \bu).
  \end{equation}
    This system has a monomial quadratization assuming that the inputs are differentiable.
   Moreover, the order of the optimal monomial quadratization does not exceed $\Pi_{i=1}^{\nn+r}(d_i + 1)$, where 
   \[
    d_i := \deg_{x_i} \bp(\bx, \bu) \;\; \text{for} \;\; 1 \leqslant i \leqslant \nn \quad\text{ and }\quad d_{\nn + i} := \deg_{u_i} \bp(\bx, \bu), \;\; \text{for} \;\; 1 \leqslant i \leqslant r,
   \]
   are the maximal degrees of the polynomials in the state and inputs, respectively. %
\end{theorem}

For systems that do not have a differentiable input, the existence of an input-free quadratization is tied to additional conditions, see~\cite[Proposition~3.7]{ByIsPoKr_Quadratization_2023}.
The following example illustrates this difference.

\begin{example}
    As we have discussed, every non-autonomous polynomial ODE system admits a quadratization (Theorem~\ref{thm:input_quadr}), and this is not true for input-free quadratizations.
    Consider the scalar ODE 
    \[
    \dot{x} = -x + x^2 u.
    \]
    First, we show that $w(x, u) = xu$ is a quadratization, since
    \[
    \dot{x} = -x + xw, \quad\quad \dot{w} = x\dot{u} + \dot{x}u = x\dot{u} + (-x + x^2 u) u = x\dot{u} - xu + w^2.
    \]
    Next we prove that the equation does not admit an input-free quadratization. 
    Assume that such quadratization $\bw(x)$ exists.
    Then the monomial $x^2u$ must be written as $\ell(x, \bw) u$, where $\ell(x, \bw)$ is a linear function in $x$ and $\bw$. 
    Then the derivative $\dot{x^2} = \dot{\ell}$ must be quadratic in $\bw, x, u$.
    Computing $\dot{x}^2 = -2x^2 + 2x^3u$, we can analogously deduce that $\dot{x}^3$ must be quadratic in $\bw, x, u$ as well.
    Then the same argument will apply to $x^4, x^5, \ldots$ leading to a contradiction.
    For a more systematic treatment of this phenomenon we refer to~\cite[Section~3.2]{ByIsPoKr_Quadratization_2023}.
\end{example}

\subsection{Overview of algorithmic results and software} \label{ss:quad:algorithms}

As we mentioned above, computing an optimal quadratization is an NP-hard problem~\cite[Theorem~2]{lifeware1} (in both monomial and general polynomial cases). Nevertheless, there are at least two practically useful and implemented algorithms for computing optimal quadratizations of low order.
The one proposed in~\cite{lifeware1}  and implemented in the {\sc BioCham} software~\cite{Biocham} takes the large quadratization built in the proof of Theorem~\ref{thm:existence_simple} and then uses SAT-solving techniques to find a smallest possible subset of it which is itself already a quadratization.
Although the optimality is not guaranteed, the produced quadratizations~\cite[Table~1]{lifeware1} are compact enough to be practically useful and often turn out to be optimal (see~\cite[Table~3]{Bychkov2021} for comparison with some optimal ones).
Another algorithm proposed in~\cite{Bychkov2021} and implemented in the {\sc QBee} Python package~\cite{QBee} takes a different approach and does not rely on the explicit quadratization given by Theorem~\ref{thm:existence_simple}.
The idea is to search the entire space of possible sets of new variables by organizing it as a tree, where the original system is the root node, and the children nodes are obtained from the parent node by adding new variables. 
While such a tree is infinite, it turns out that one can guarantee that the optimal quadratization will always be located in a certain finite subtree; this is explored efficiently by the algorithm.
The algorithm and implementation were recently extended in~\cite{ByIsPoKr_Quadratization_2023} to compute quadratizations for non-autonomous systems in the sense of Definitions~\ref{def:quadr_input} and~\ref{def:quadr_input_zero}. Furthermore, an extension of {\sc QBee} was proposed in~\cite{Yubo} to compute quadratizations that preserve dissipative equilibria.

\section{Applications and Examples} \label{sec:application}
We focus on two examples from different communities: one from neural network approximation and one from cell signaling. We refer the reader to other work where nontrivial lifting transformations have been discovered, e.g., in chemical reaction networks~\cite{lifeware1}, solar wind modeling~\cite{ByIsPoKr_Quadratization_2023}, rocket combustion~\cite{SKHW2020_learning_ROMs_combustor}, compressible Euler equations and a continuously variable resonance combustor ~\cite{QKMW2019_transform_and_learn}, and additive manufacturing~\cite{khodabakhshi2022non}.

\subsection{Quadratizing a sigmoid perceptron} \label{ss:sigmoid}
A popular approach to data-driven modeling of dynamical systems of the form $\dot{\bx} = \bbf(\bx)$ is to fit $\bbf$ from the data using, for example, a neural network.
We consider one of the simplest nontrivial cases of a one-dimensional dynamical system that is approximated using a single-layer perceptron with a sigmoid activation function. The resulting scalar ODE is 
\begin{equation}\label{eq:sigmoid}
\dot{x} = \frac{1}{1 + e^{-ax - b}},
\end{equation}
where $a, b \in \real$ are the weight and the offset, respectively.
Performing polynomialization and quadratization with {\sc QBee}, we obtain the new variables 
\[
w_1 = \frac{1}{1 + e^{-ax - b}}, \qquad w_2 = e^{-ax}, \qquad w_3 = w_1^2 w_2 = \frac{e^{-ax}}{(1 + e^{-ax - b})^2},
\]
and the quadratic system
\[
\begin{cases}
    \dot{x} = w_1,\\
    \dot{w}_1 = a e^{-b} w_1 w_3,\\
    \dot{w}_2 = -a w_1 w_2,\\
    \dot{w}_3 = 2a e^{-b} w_3^2 - a w_1 w_3.
\end{cases}
\]
While the system above has dimension four, we note that the variables $w_1$ and $w_3$ form a closed subsystem
\[
\begin{cases}
    \dot{w}_1 = a e^{-b} w_1 w_3,\\
    \dot{w}_3 = 2a e^{-b} w_3^2 - a w_1 w_3.
\end{cases}
\]
This system can be further simplified by setting $w_4 = e^{-b}w_3$ into
\begin{equation}\label{eq:sigmoid_final}
\begin{cases}
    \dot{w}_1 = a w_1 w_4,\\
    \dot{w}_4 = 2a w_4^2 - a w_1 w_4.
\end{cases}
\end{equation}
From a trajectory of~\eqref{eq:sigmoid_final}, one can reconstruct $x$ as $x = \frac{-1}{a} \ln\left(\frac{w_4}{w_1^2} \right) - \frac{b}{a}$.

To recap, we transformed the original scalar non-polynomial model with two parameters to a two-dimensional quadratic model with only one parameter.
In particular, although we added new variables, the degrees of freedom of the model remain the same: any trajectory of~\eqref{eq:sigmoid} was uniquely defined by $(x(0), a, b)$, and any trajectory of~\eqref{eq:sigmoid_final} is uniquely defined by $(w_1(0), w_4(0), a)$. However, note that the interpretations of the degrees of freedom have changed. While the former had one state and two parameters, the quadratized system has two states and one parameter. 

While the first step of the transformation yielding a four-dimensional model was fully algorithmic, the last step that rescaled by $e^{-b}$ can also be performed by the algorithms from~\cite{Hubert2013}. However, the second step that restricted the model to a subsystem of two dimensions was performed manually (a similar trick was employed in~\cite[Section~7]{ByIsPoKr_Quadratization_2023}).
It is an interesting future research problem to automatize this step as well.

Interestingly, this two-dimensional system~\eqref{eq:sigmoid_final} is equivalent to the following elementary chemical reaction network (CRN) (see~\cite{lifeware1} for definitions):
\[
W_1 + W_4 \xrightarrow{a} 2W_1, \quad W_1 \xrightarrow{1} W_1 + 2W_4.
\]
CRNs are at the core of arbitrary-precision Turing-complete analog computing, and a famous result states that any real computable function can be approximated by a CRN with a finite number of species~\cite{bournez2007polynomial}.
Bimolecular reactions, which are represented by quadratic ODEs, are particularly appealing from an experimental perspective~\cite{lifeware1}. Thus, (optimal) quadratization provides an important step in analog computing by reducing the amount of molecular species (the number of variables) required in the analog computer.

\subsection{MAPK model}
We consider a representative phenomenological model of cell signaling, the mitogen-activated protein kinase (MAPK) pathway from
\cite[Eq. (9)]{Nguyen2015}; here, we follow the notation of~\cite[Eq. (32)]{Linden2022}. The model includes mixed feedback (negative and positive feedback), which is necessary to predict the range of dynamical behavior observed in experiments. The model's three states, $x_1(t), x_2(t), x_3(t)$ are the phosphorylated RAF, MEK, and MAPK/ERK, and they evolve as
\begin{equation}\label{eq:mapk}
   \begin{cases}
       \dot{x}_1 = k_1 (S_{1t} - x_1) \frac{K_1^{n_1}}{K_1^{n_1} + x_3^{n_1}} - k_2 x_1,\\
       \dot{x}_2 = k_3 (S_{2t} - x_2) x_1 \left( 1 + \frac{\alpha x_3^{n_2}}{K_2^{n_2} + x_3^{n_2}}\right) - k_4 x_2,\\
       \dot{x}_3 = k_5 (S_{3t} - x_3) x_2 - k_6 x_3,
   \end{cases} 
\end{equation}
where $k_1, \ldots, k_6, K_1, K_2, n_1, n_2, S_{1t}, S_{2t}, S_{3t}, \alpha$ are scalar parameters.
Computation with {\sc QBee} shows that~\eqref{eq:mapk} can be quadratized using 15 new variables:
\begin{align*}
    w_1 &= x_3^{n_1},\;\; &w_2 &= x_3^{n2},\;\; &w_3 &= \frac{1}{K_1^{n_1} + x_3^{n_1}},\;\; &w_4 &= \frac{1}{K_2^{n_2} + x_3^{n_2}},\\
    w_5 &= \frac{x_1}{x_3},\;\; &w_6 &= \frac{x_2}{x_3},\;\; &w_7 &= \frac{x_3^{n_1}}{K_1^{n_1} + x_3^{n_1}},\;\; &w_8 &= \frac{x_3^{n_2}}{K_2^{n_2} + x_3^{n_2}},\\
    w_9 &= \frac{w_3}{x_3},\;\; &w_{10} &= w_7 x_2,\;\; &w_{11} &= w_8 x_2,\;\; &w_{12} &= w_8 x_1,\\
    w_{13} & = w_6 w_7,\;\; &w_{14} &= w_6 w_8,\;\;  &w_{15} &= w_5 w_8. & &
\end{align*}
We omit the resulting quadratic state equations for space reasons.
Obtaining a quadratization for a system of this complexity by hand would be an extremely challenging task, so having automatic software and algorithms allows for the discovery of completely new quadratizations.
However, as mentioned above (e.g., in Example~\ref{ex:poly_nonoptimal}), the existing polynomialization algorithms do not have optimality guarantees: here, one could notice that $w_3 = \frac{1 - w_7}{K_1^{n_1}}$ and $w_4 = \frac{1 - w_8}{K_2^{n_2}}$ are linear in the other variables, so they can be removed from the set of new variables without losing the quadratization property.
This results in a quadratic system with $13$ quadratization variables, two less than found by {\sc QBee}.

This opens up a very interesting scientific question from a systems biology perspective: It would be interesting to consider the elementary chemical reaction network arising from this quadratized system. One could then compare it with the corresponding networks arising in cells, similar to the chemical reaction network interpretation of the Hill activation function studied in~\cite[Example~1]{lifeware1}.

\section{Open problems} \label{sec:problems}
There are several open problems regarding the theory and practical implementation of polynomialization and quadratization that we think would be of great interest to the scientific community. These---possibly incomplete---research questions are as follows. 

\begin{itemize}
    \item While existing quadratization algorithms such as {\sc QBee} provide some optimality guarantees (namely, producing optimal monomial quadratization), this is not the case for the polynomialization problem.
    As Example~\ref{ex:poly_nonoptimal} demonstrates, the suboptimality of the existing approaches can already be seen for simple scalar ODEs.
    Providing optimality guarantees for polynomialization, at least with restricted type of nonlinearities (e.g., only exponential and trigonometric functions), is a challenging problem both on the theoretical and practical level.
    
    \item The general theory presented in Theorem~\ref{thm:polynomialization} establishes the possibility of polynomialization for differential algebraic equations (DAEs).
    In contrast, the existing algorithmic tools allow only expressions in several standard nonpolynomial functions (such as exponential, trigonometric, and logarithmic) omitting many important special functions, such as the Bessel function in Example~\ref{ex:bessel}).
    Based on the discussion in Example~\ref{ex:bessel}, a way to fill this gap could be to allow the addition of new special functions on the fly by providing a differential equation satisfied by the function.
    Adding such a feature to existing algorithms is an important challenge from both a theoretical and also software design standpoint.

    \item Once polynomialization and quadratization become available thanks to recent advances in theory and software, the next natural question is to search for lifting transformations that preserve important properties of the original system such as stability, which can be characterized for quadratic ODEs as in~\cite{K2020_stability_domains_QBROMs}.
    In the context of quadratization, recent results allow preserving dissipativity~\cite{Yubo} and ensuring stability of periodic orbits~\cite{Bayer2024} while more general problems related to global stability remain open.
    For the polynomialization problem, all these questions are open.

    \item Existing quadratization tools search for new variables among monomials (that is, products of positive powers of variables such as $x_1 x_2^3$).
    It has been observed that one can obtain quadratizations of smaller dimension if we allow negative powers~\cite[Proposition~1]{Bychkov2021} or arbitrary polynomials~\cite{Alauddin}.
    For the moment, none of these insights have been used systematically in general algorithms, and developing such new algorithms with more expressive language for quadratization is an important open problem.

    \item This survey focuses on lifting ODEs to ODEs.
    However, it has been recognized (e.g., \cite{guillot2019generic,KW18nonlinearMORliftingPOD,khodabakhshi2022non}) that it may be easier to lift an ODE to a polynomial or quadratic DAE, and this lifting is sufficient for some applications.
    It would be great to have solid foundations and automatic software tools for such transformations.

    \item Finally, an important question is generalizing the quadratization and polynomialization algorithms to partial differential equations.
    This would allow the new sets of variables to describe complex dynamics to be leveraged in operator learning, for Koopman observables, for reachability analysis, and potentially as structured nonlinear manifolds to uncover new latent space representations. 
    Promising preliminary results that rely on ODE quadratization with {\sc QBee} have been obtained in~\cite{ByIsPoKr_Quadratization_2023}.
\end{itemize}


\section{Conclusions} 
\label{sec:conclusions}
We presented an introduction and overview of methods of polynomialization and quadratization in the context of nonpolynomial and higher-degree polynomial finite-dimensional dynamical systems. These two processes have been studied and applied in a variety of disciplines for more than a century, with little interaction between the developers of the theory and the practitioners that can benefit from and interpret scientifically the resulting systems. We hope that this introductory article can contribute to a broader interest in the methods and spur further research in theory, algorithms, and software for quadratization and polynomialization.

\subsubsection*{Acknowledgments}
B.K. is supported in part through NSF-CMMMI award 2144023. 
G.P. was partially supported by the INS2I PANTOMIME project and the French ANR-22-CE48-0016 NODE project.

%
%
\bibliographystyle{splncs04}
\bibliography{references}
\end{document}

%% file: mymacros.tex
\newcommand{\bit}{\begin{itemize}}
\newcommand{\eit}{\end{itemize}}
\newcommand{\ben}{\begin{enumerate}}
\newcommand{\een}{\end{enumerate}}

\newcommand {\KK} {\mathbb{K}}
\newcommand {\real} {\mathbb{R}}

\newcommand {\compl} {\mathbb{C}}

\newcommand{\bbf}{\ensuremath{\mathbf{f}}}

\newcommand{\bp}{\ensuremath{\mathbf{p}}}
\newcommand{\bq}{\ensuremath{\mathbf{q}}}

\newcommand{\bu}{\ensuremath{\mathbf{u}}}

\newcommand{\bw}{\ensuremath{\mathbf{w}}}
\newcommand{\bx}{\ensuremath{\mathbf{x}}}

